\documentclass[conference]{IEEEtran}
\IEEEoverridecommandlockouts
\usepackage{cite}
\usepackage{amsmath,amssymb,amsfonts}
\usepackage{algorithmic}
\usepackage{graphicx}
\usepackage{textcomp}
\usepackage{xcolor}
\usepackage{float}
\usepackage[caption = false]{subfig}
\usepackage{siunitx}
\usepackage{scalerel}
\usepackage{multirow}
\usepackage{pgfplots}
\usepackage{soul}
\usepackage{acronym}
\pgfplotsset{compat=newest}
\usepackage[center]{caption}
\usepackage{eurosym}
\usepackage{multicol}
\usepackage{booktabs}

\DeclareSIUnit{\var}{Var}
\DeclareSIUnit{\va}{VA}
\DeclareSIUnit{\pu}{p.u.}
\DeclareSIUnit{\deg}{deg}
\DeclareSIUnit{\MW}{MW}
\DeclareSIUnit{\MWh}{MWh}
\DeclareSIUnit{\min}{min}
\DeclareSIUnit{\t}{t}
\DeclareSIUnit{\EUR}{\text{\euro}}

\acrodef{bess}[BESS]{Battery Energy Storage System}
\acrodef{eedi}[EEDI]{Energy Efficiency Design Index}
\acrodef{imo}[IMO]{International Maritime Organization}
\acrodef{epla}[EPLA]{Electric Power Load Analysis}
\acrodef{dg}[DG]{Diesel Generator}
\acrodef{seemp}[SEEMP]{Ship Energy Efficiency Management Plan}
\acrodef{cii}[CII]{Carbon Intensity Indicator}
\acrodef{uc}[UC]{Unit Commitment}
\acrodef{milp}[MILP]{Mixed Integer Linear Programming}
\acrodef{ghg}[GHG]{Global Greenhouse Gases}
\acrodef{nox}[$NO_X$]{Nitrogen Oxide}
\acrodef{sox}[$SO_X$]{Sulphur Oxide}
\acrodef{pm}[PM]{Particulate Matter}
\acrodef{un}[UN]{United Nation}
\acrodef{vlsfo}[VLSFO]{Very Low Sulphur Fuel Oil}
\acrodef{egcs}[EGCS]{Exhaust Gas Cleaning Systems}
\acrodef{eedi}[EEDI]{Energy Efficiency Design Index}
\acrodef{eexi}[EEXI]{ Energy Efficiency Existing Ship Index}
\acrodef{cii}[CII]{Carbon Intensity Indicator}
\acrodef{gt}[GT]{Gross Tonnage}
\acrodef{mc}[MC]{Markov Chain}
\acrodef{sfoc}[SFOC]{Specific Fuel Oil Consumption}
\acrodef{soc}[SOC]{State Of Charge}
\acrodef{sog}[SOG]{Speed Over Ground}
\acrodef{oc}[OC]{Operating Condition}
\acrodef{igsc}[IGSC]{Instantaneous Growing Stream Clustering}
\acrodef{swbs}[SWBS]{Ship Work Breakdown Structure}
\acrodef{sps}[SPS]{Shipboard Power System}
\acrodef{mepc}[MEPC]{Marine Environment Protection Committee}
\acrodef{dcs}[DCS]{Data Collection System}
\acrodef{eeoi}[EEOI]{Energy Efficiency Operational Indicator}
\acrodef{solas}[SOLAS]{Safety Of Life At Sea}
\acrodef{sgp}[SGP]{State Growth Parameter}
\acrodef{eca}[ECA]{Emission Control Area}
\acrodef{aes}[AES]{All Electric Ship}
\acrodef{pms}[PMS]{Power Management System}
\acrodef{iacs}[IACS]{International Association of Classification Societies}

\begin{document}

\title{A Security-Constrained Optimal Power Management Algorithm for Shipboard Microgrids with Battery Energy Storage System}

\author{\IEEEauthorblockN{F. D'Agostino, M. Gallo, M. Saviozzi, F. Silvestro}
\IEEEauthorblockA{\textit{University of Genova}\\
    \textit{DITEN - Dipartimento di Ingegneria Navale, Elettrica, Elettronica e delle Telecomunicazioni}\\
    Genova, Italy\\
    fabio.dagostino@unige.it, marco.gallo@edu.unige.it, matteo.saviozzi@unige.it, federico.silvestro@unige.it}
}

%%%%%%%%%%%%%%%%% IEEE COPYRIGHT %%%%%%%%%%%%%%%%%
\makeatletter

\def\ps@IEEEtitlepagestyle{%
  \def\@oddfoot{\mycopyrightnotice}%
  \def\@evenfoot{}%
}
\def\mycopyrightnotice{%
  {\footnotesize 979-8-3503-4689-3/23/\$31.00~\copyright2023 IEEE\hfill}%
  \gdef\mycopyrightnotice{}
}

\maketitle

\begin{abstract}
This work proposes an optimal power management strategy for shipboard microgrids equipped with diesel generators and a battery energy storage system. The optimization provides both the unit commitment and the optimal power dispatch of all the resources, in order to ensure reliable power supply at minimum cost and with minimum environmental impact. The optimization is performed solving a mixed integer linear programming problem, where the constraints are defined according to the operational limits of the resources when a contingency occurs. The algorithm is tested on a notional all-electric ship where the ship's electrical load is generated through a Markov chain, modeled on real measurement data. The results show that the proposed power management strategy successfully maximizes fuel saving while ensuring blackout prevention capability.
\end{abstract}


\section{Introduction} \label{Intro}
Ecological transition is one of the main topics addressed nowadays. Emissions from maritime transport accounts for about 3\% of \ac{ghg}, as well as 13\% of \ac{nox} and 12\% of \ac{sox} emissions, including \ac{pm}, methane, all known to be harmful to human health \cite{MUELLER2023114460}.

In order to limit the emission of the ships, in 1973, the \ac{imo} adopted the International Convention for the Prevention of Pollution from Ships known as MARPOL \cite{MARPOL}. From 2020 the use of \ac{vlsfo} has become mandatory (0.50\% sulphur limit).
This is in agreement to the "European Union’s Fit for 55" climate package of legislative proposals \cite{fit455}. 
These policies include measures to reduce \ac{ghg} emissions by 55\% by 2030 compared to 1990 values.

Several recent papers propose the utilization of \ac{bess} as a mean to improve efficiency in \ac{aes} where the electrical generating is provided by \ac{dg}. 
In \cite{BESSuse}, a wide range of functions for \ac{bess} is described. 
The strategic loading is an interesting function wherein the \ac{bess} is exploited to optimize the working point of the \ac{dg}.

In \ac{aes} one of the main challenges is to design the \ac{pms} strategy that coordinates the power sources to achieve efficient and robust operation.
As per the \ac{iacs} guidelines, in the event of a failure of one generating unit, the system must be able to avoid the blackout \cite{IACS} . Therefore, the security constraints need to take into account the two main limitations of each generating unit: the maximum overload and the maximum permissible load step that a generator can absorb.

There are different strategies that have been presented in literature. 
In \cite{RadanBlackout}, a rules based power management strategy is proposed with the aim of increasing the anti blackout capability while minimising fuel consumption. An optimization is carried out to realize a load dependent start-up table, which rules the status of the generators (on, off).
In \cite{twostepPMS}, a two step multi-objective optimization method for \ac{aes} is proposed. 
In \cite{SecurityUC}, a security constrained power management strategy is designed to optimally operate the system and to guarantee its security. 
In \cite{kanellos}, an optimal power management method is proposed so that the operational costs are minimized.

In this paper an optimal power management strategy for shipboard microgrids equipped with \acp{dg} and a \ac{bess} is proposed. 
The algorithm, based on an optimization problem, provides both the \ac{uc} and the economic dispatch of all generating units.
The proposed methodology is based on a \ac{milp} problem wherein security constraints are modelled in order to avoid the blackout in the event of a failure. 
The modelling of the security constraints represents the main contribution of this paper.
The algorithm is validated through simulations on a notional cruise ship with four main \acp{dg} and a \ac{bess} where the electrical load is modelled starting from real measurements. 

The rest of the paper is organized as follows: Section II introduces the System Modelling adopted for the \ac{pms}, Section III provides the Optimization Problem, Section IV reports Simulation and Results Analysis while the conclusions are reported in Section V.

\section{System Modelling}
\begin{figure}[H]
	\centering
	\includegraphics[trim={0cm 2cm 0cm 2cm}, width=\columnwidth]{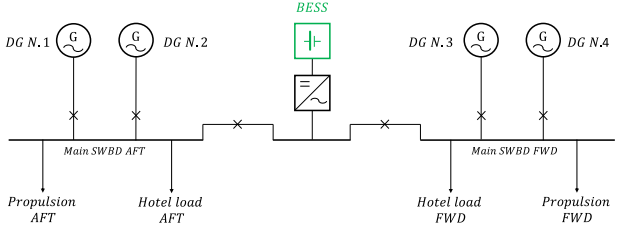}
	\caption{System Architecture.}\label{fig:SystemArchitecture}
\end{figure}

Figure~\ref{fig:SystemArchitecture} reports the notional architecture of the selected cruise ship. The generating resources of the ship are composed of \acp{dg} and a \ac{bess} that are based on real components.

In an \ac{aes} the main load is represented by the propulsion system, which is a function of the speed of the ship. The extra propulsive ship's load (e.g. HVAC, galley, accommodation, etc.) is represented by the hotel load \cite{EPLAiees}.
Each of these loads are evaluated in the \ac{epla} of the ship that are divided according to the \acp{oc} (e.g. full power, navigation, etc.) of the ship \cite{Doerry}.

\subsection{\ac{dg} Fuel Consumption} \label{DGmodel}
\begin{figure}[H]
	\centering
        \includegraphics[width=\columnwidth]{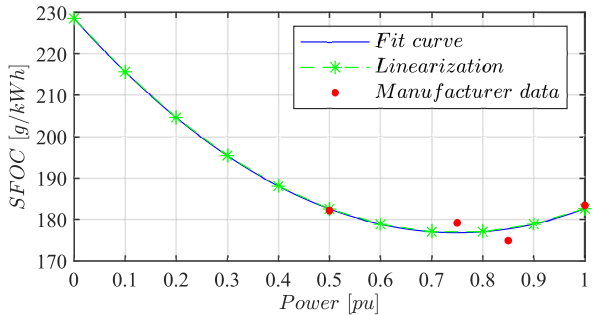}
	\caption{\ac{sfoc} curve.}\label{fig:SFOCcurve}
\end{figure}

The \ac{dg} fuel consumption is modelled by a linearized \ac{sfoc} curve reported in Fig.~\ref{fig:SFOCcurve}. 
The linearization is performed in two steps: (\textit{i}) fitting the real data, (\textit{ii}) linearize the fitted curve. 

There is a non-linear relationship between \ac{sfoc} and \ac{dg}'s power. From the manufacturers data-sheet the \ac{sfoc} is obtained at some working point of the power of the \ac{dg} \cite{wartsila46f}. These data are fitted by a polynomial regression. As in \cite{SFOCparbolic}, it is assumed a parabolic function between \ac{sfoc} and the power of the \ac{dg}. From this regression a linearization is then performed. 

Figure~\ref{fig:SFOCcurve} shows the results of the linearization that divides the quadratic function in 10 equal intervals.

The $m$-th interval is a line with slope $a_{m,i}(t)$ and intercept $b_{m,i}(t)$ ($i$ indicates the $i$-th \ac{dg}).

\subsection{Battery Energy Storage System}
The \ac{soc} of the battery is modelled according to the following equations as in \cite{SOCdef}: 
\begin{equation}
    SOC(t+1) = 
    \begin{aligned}[t]
    & SOC(t) + \left(P_{b}^{d}(t) \cdot \frac{1}{\eta_{d}} + \right.\\
    & - P_{b}^{c}(t) \cdot \eta_{c} \biggl) \cdot \frac{\Delta t}{E_n} \quad \forall \; t = 1:T\label{eq:SOCbatteria}
    \end{aligned}
\end{equation}

Where $E_n$ [\SI{}{\kWh}] is the energy of the battery, $P_{b}^{c}$ [\SI{}{\kW}] and $P_{b}^{d}$ [\SI{}{\kW}] are respectively the charge and discharge power of the battery, $\Delta t$ is the granularity of the control and $T$ is the horizon of the optimization. 
The \ac{bess} is composed by an energy storage and a conversion system. These components are modelled through the efficiency of the \ac{bess} which is different for the charging ($\eta_{c}$) and the discharging ($\eta_{d}$) of the storage system.
It is assumed that the initial ($t = 1$) \ac{soc} is known and it is equal to the final ($t = T$) value.

\section{Optimization Problem}
The proposed methodology is based on a \ac{milp} optimization algorithm implemented in MATLAB environment.
Several constraints are implemented in the algorithm to model the \acp{dg}, the \ac{bess} and to manage the security of the system.

\subsection{\ac{dg} Constraints}
The linearization of the \ac{sfoc} curve of the \acp{dg} is modelled by the equation as in \cite{PIECEWISELIN}:
\begin{align}
    & P_{DG,i}(t) = \sum\limits_{m=1}^{n_i} \delta_{m,i}(t) \quad \forall \; i = 1:N, \forall \; t = 1:T\label{eq:PieceWiseLin1}\\
    & 0 \le \delta_{m,i}(t) \le \frac{P_{DG,i}^{n}}{n_i} \quad \forall \; i = 1:N, \forall \; t = 1:T\label{eq:PieceWiseLin2}
\end{align}
where $P_{DG,i}(t)$ is the active power set-point of the $i$-th \ac{dg}, $\delta_{m,i}(t)$ is the auxiliary variable of the $m$-th interval of the $i$-th \ac{dg}, $P_{DG,i}^{n}$ is the rated power of the $i$-the \ac{dg}, $N$ is the total number of \acp{dg}, and $n_i$ indicates the number of linearization intervals. 

The active power of the \ac{dg}s is bounded between a maximum \eqref{eq:P_DGmax} and a minimum \eqref{eq:P_DGmin} set-point:
\begin{equation}
    P_{DG,i}(t) \le c_{DG,max} \cdot P_{DG,i}^{n} \cdot 
    \begin{aligned}[t]
    & z_i(t) \\
    & \forall \; i = 1:N, \forall \; t = 1:T\label{eq:P_DGmax}
    \end{aligned}
\end{equation}
\begin{equation}
    P_{DG,i}(t) \ge c_{DG,min} \cdot P_{DG,i}^{n} \cdot 
    \begin{aligned}[t]
    & z_i(t) \\
    & \forall \; i = 1:N, \forall \; t = 1:T\label{eq:P_DGmin}
    \end{aligned}
\end{equation}

In these equations the parameters $c_{DG,min}$ and $c_{DG,max}$ are respectively the coefficients that allow to evaluate the minimum and the maximum active power of the $i$-th \ac{dg} starting from $P_{DG,i}^{n}$.
The binary variable $z_i(t)$ identifies the status of the $i$-th \ac{dg} ($z_i(t) = 1$ means that the $i$-th \ac{dg} is ON at time $t$).
The equation \eqref{eq:StartUpConst} models the start-up of the \acp{dg}. 
\begin{align}
    & z_i(t)-z_i(t-1) \le u_i(t) \quad \forall \; i = 1:N, \forall \; t = 1:T\label{eq:StartUpConst}
\end{align}
The binary variable $u_i(t)$ is equal to 1 if the $i$-th \ac{dg} is starting up at time $t$.

The equations \eqref{eq:MinUpTime}--\eqref{eq:MinDownTime} models the minimum up and the minimum down times of the \acp{dg}.
\begin{equation}
    u_i(t) - \frac{\Delta t}{t_{min_{u},i}} \sum\limits_{u_t=1}^{t_{min_{u},i}/\Delta t}
    \begin{aligned}[t]
    & {z_i\left(t + u_t \right)} \le 0 \\
    & \forall \; i = 1:N, \forall \; t = 1:T\label{eq:MinUpTime}\\
    \end{aligned}
\end{equation}
\begin{equation}
    u_i(t) + \frac{\Delta t}{t_{min_{d},i}} \sum\limits_{d_t=1}^{t_{min_{d},i}/\Delta t}
    \begin{aligned}[t]
    & {z_i\left(t - d_t \right)} \le 1 \\
    & \forall \; i = 1:N, \forall \; t = 1:T\label{eq:MinDownTime}\\
    \end{aligned}
\end{equation}

Where $t_{min_{d},i}$ and $t_{min_{u},i}$ are respectively the minimum down time and the minimum up time of the $i$-th \ac{dg}. Both of these parameters must be multiples of the simulation time-step.
Equations \eqref{eq:MaxRamp}--\eqref{eq:MinRamp} represent the ramp limits of the \acp{dg}. 
\begin{equation}
    P_{DG,i}(t) - P_{DG,i}(t-1)
    \begin{aligned}[t]
    & \le \Delta P_{r_u,i} \\
    & \forall \; i = 1:N, \forall \; t = 1:T\label{eq:MaxRamp}\\
    \end{aligned}
\end{equation}
\begin{equation}
    P_{DG,i}(t-1) - P_{DG,i}(t)
    \begin{aligned}[t]
    & \le \Delta P_{r_{d,i}} \\
    & \forall \; i = 1:N, \forall \; t = 1:T\label{eq:MinRamp}\\
    \end{aligned}
\end{equation}

Where $\Delta P_{r_u,i}$ and $\Delta P_{r_d,i}$ are respectively the maximum power ramp-up and ramp-down limit that a \ac{dg} can take in a single time-step.

\subsection{\ac{bess} Constraints}
The \ac{soc} of the \ac{bess} is limited between a minimum ($SOC_{min}$) and maximum ($SOC_{max}$) value according to the following constraints:
\begin{align}
    & SOC_{min} \le SOC(t) \le  SOC_{max} \quad \forall \; t = 1:T \label{eq:SOCconstr}
\end{align}

Equations \eqref{eq:PBESSconstrChargMax} -- \eqref{eq:PBESSconstrDischargMin} represent the upper and lower bound of the charging power ($P_{b}^{c}(t)$) and discharging power ($P_{b}^{d}(t)$) of the battery at time $t$. 
\begin{align}
    & P_{b}^{c}(t) \le c_{c,max} \cdot P_{b}^{n} \cdot z_b^c(t) \quad \forall \; t = 1:T \label{eq:PBESSconstrChargMax} \\
    & P_{b}^{c}(t) \ge c_{c,min} \cdot P_{b}^{n} \cdot z_b^c(t) \quad \forall \; t = 1:T \label{eq:PBESSconstrChargMin} \\
    & P_{b}^{d}(t) \le c_{d,max} \cdot P_{b}^{n} \cdot z_b^d(t) \quad \forall \; t = 1:T \label{eq:PBESSconstrDischargMax} \\
    & P_{b}^{d}(t) \ge c_{d,min} \cdot P_{b}^{n} \cdot z_b^d(t) \quad \forall \; t = 1:T \label{eq:PBESSconstrDischargMin}
\end{align}

In these equations $c_{c,max}$,$c_{c,min}$,$c_{d,max}$ and $c_{d,min}$ represent respectively the max/min charge and the  max/min discharge of the battery, while $z_b^c(t)$/$z_b^d(t)$ are two binary variable that identify if the \ac{bess} is charging/discharging at time $t$.
Equation \eqref{eq:ChargeOrDisch} models that the battery can only charge or discharge in a single time-step.
\begin{align}
    & z_b^d(t) + z_b^c(t) = z_b(t) \quad \forall \; t = 1:T\label{eq:ChargeOrDisch}
\end{align}
Where $z_b(t)$ is a binary variable that is equal to 1 if the \ac{bess} is charging or discharging a time $t$.

\subsection{Security Constraints}
As introduced in Section~\ref{Intro}, the \ac{iacs} requires that if one of the generating units fail, the remaining units must be able to avoid blackout. This means that at least 2 generating units must be in service at all time instants $t$ to meet the electrical load. 
Auxiliary variables are implemented in order to model the security constraints. First of all, it is necessary to identify which of the $n_g$ generating units (\acp{dg} and \ac{bess}) are on. 

There are $n_c$ combinations without repetition of class $k$ greater or equal to 2 of generating units that are providing power. For instance, if the number of total units are 3 there are 4 possible combinations of at least 2 units that are in service.
It is possible to calculate this number using the binomial coefficient $C(n_g,k)$ for each $k$ class and then sum all the coefficient where $n_g = N + 1$. 

Thus, in the above mentioned example, since at least 2 units must be ON, the combinations are:
\begin{equation}
    \begin{aligned}[t]
    & n_c = \sum\limits_{k \ge 2} C(n_g,k) = \binom{3}{2} + \binom{3}{3} = \\
    & = \frac{3!}{2!(3-2)!} + \frac{3!}{3!(3-3)!} = 4\label{eq:Combinations}
    \end{aligned}
\end{equation}

The first binary auxiliary variable is $s_j(t)$. If the $j$-th combination of generators is active then $s_j(t)=1$.
Since at time $t$ only one combination can be active, the following equations models the fact that only one combination can be active:
\begin{align}
    & v(t) \cdot \left(1 - \sum\limits_{j=1}^{n_c} s_j(t)\right)  = 0 \quad \forall \; t = 1:T\label{eq:SecConstr1}
\end{align}

In \eqref{eq:SecConstr1}, $v(t)$ is a parameter that can be 1 if the constraints is active at time $t$, and 0 otherwise.
The activation of the constraint depends on the \ac{oc} of the ship as it will be seen in Section~\ref{SimAndRes}. 

The second auxiliary variable is $f_j(t)$. In \eqref{eq:SecConstr2}, it is modelled the number of units that are supplying power for each $j$-th combination.
\begin{equation}
    v(t) \cdot \left(f_j(t)- \sum\limits_{r=1}^{n_j} z_{gen,r}(t)\right)
    \begin{aligned}[t]
    & = 0  \\ 
    & \forall \; j = 1:n_c, \forall \; t = 1:T\label{eq:SecConstr2}
    \end{aligned}
\end{equation}

Where the binary variable $z_{gen,r}(t)$ is equal to 1 if $r$-th unit is on at time $t$ and $n_j$ represents the possible generating unit of the $j$-th combination.

The inequality \eqref{eq:SecConstr3} links the variable $s_j(t)$ with $k_j(t)$ ensuring that when $s_j(t) = 1$ then $f_j(t) = n_j$.
\begin{equation}
    v(t) \cdot \biggl(f_j(t) - n_j \cdot s_j(t)\biggl)
    \begin{aligned}[t]
    & \ge 0  \\ 
    & \forall \; j = 1:n_c, \forall \; t = 1:T\label{eq:SecConstr3}
    \end{aligned}
\end{equation}

The following constraint ensures that, following the loss of a generating unit, the remaining $n_j - 1$ are able to provide the total electric power load:
\begin{equation}
    v(t) \cdot \biggl(s_j(t)
    \begin{aligned}[t]
    & \cdot P_{load}(t) + P_{b}^{c}(t) +\\
    & - \sum\limits_{h=1,h \ne r}^{n_j - 1} \alpha_{gen,h}  \cdot P_{gen,h}^{n} \cdot z_{gen,h}(t) \biggl) \le 0 \\ 
    & \forall \; j = 1:n_c, \forall \; r = 1:n_j, \forall \; t = 1:T\label{eq:SecConstr4}
    \end{aligned}
\end{equation}
where $P_{load}(t)$ is the total power load of the ship at time $t$, $\alpha_{gen,h}$ represents the overload working limit that the $h$-th generator can guarantee in the event of an emergency for a short period.

The equation \eqref{eq:SecConstr5} models that a generating unit can provide a maximum instantaneous load step in case of emergency. 
\begin{equation}
    v(t) \cdot \biggl(
    \begin{aligned}[t]
    & P_{gen,r}(t) - \sum\limits_{h = 1,h \ne r}^{n_j - 1} \beta_{gen,w} \cdot P_{gen,w}^{n}  + \\
    & - M \cdot (1-s_j(t))\biggl) \le 0 \\
    & \forall \; j = 1:n_c, \forall \; r = 1:n_j, \forall \; t = 1:T\label{eq:SecConstr5}
    \end{aligned}
\end{equation}

Where $\beta_{gen,w}$ represents the instantaneous variation of power of the, the variable $P_{gen,r}(t)$ represents the power supplied by the $r$-th unit at time $t$.

Finally, equation \eqref{eq:SecConstr6} takes into account that in case of an emergency each unit is able to provide the maximum load step.
\begin{equation}
    v(t) \cdot \biggl(
    \begin{aligned}[t]
    & P_{gen,r}(t) - (\alpha_{gen,r} - \beta_{gen,r})\cdot P_{gen,r}^{n} + \\
    & - M \cdot (1-s_j(t))\biggl) \le 0 \\
    & \forall \; j = 1:n_c, \forall \; r = 1:n_j, \forall \; t = 1:T\label{eq:SecConstr6}
    \end{aligned}
\end{equation}

These two last inequalities are characterized there is the $M$ parameter. 
It model the so called Big M method in order to modify the constraints according to the activation of the $j$-th combination. In fact, when $s_j(t) = 0$ the constraint is always satisfied since the term $-M \cdot (1-s_j(t))$ is negative and it is dominants to the other contribution (e.g. $M = 10^9$).

\subsection{Load balance}
The generated electric power must be equal to the load demand that includes the \ac{bess} when it is absorbing power. This is modelled in the equation \eqref{eq:DemandGeneration} where $P_{load}$ is the total electric power require by the ship.
\begin{equation}
    P_{load}(t) + P_{b}^c(t) =  \sum\limits_{i=1}^{N} P_{DG,i}(t) 
    \begin{aligned}[t]
    & + P_{b}^d(t) \\
    & \forall \; t = 1:T\label{eq:DemandGeneration}
    \end{aligned}
\end{equation}

\subsection{Objective Function}
The objective function is formulated as:
\begin{equation}
    \min \sum\limits_{t=1}^{T} \sum\limits_{i=1}^{N} 
    \begin{aligned}[t]
    & \left( \sum\limits_{m=1}^{n_i} a_{m,i}(t) \cdot \delta_{m,i}(t) \cdot c_{fuel} + \right.\\
    & + b_{1,i}(t) \cdot z_i(t) \cdot c_{f} + c_i \cdot u_i(t) \biggl) \cdot \Delta t\label{eq:ObjFcn}
    \end{aligned}
\end{equation}

Where $c_{f}$ is the cost of the fuel [\SI[per-mode=symbol]{}{\EUR\per\kg}] and $c_i$ represents the start-up cost of the $i$-th generator.
The optimization algorithm minimizes the total cost [\SI{}{\EUR}] that is divided in two parts: one is the fuel cost and the other is the start-up cost.

Summarizing the optimization problem is composed of the objective function \eqref{eq:ObjFcn} and the constraints \eqref{eq:PieceWiseLin1}-\eqref{eq:SecConstr6}. 
In order to facilitate the paper comprehension, Table~\ref{tab:OptiVar} reports the variables of the optimization problem.
\begin{table}[h]
		\centering
		\caption{Variables of the optimisation problems.}
		\label{tab:OptiVar}
		\renewcommand{\arraystretch}{1.4}
        \begin{tabular}{ c l  l  }
        \hline \hline
        Variable                       & Description                                                                                                                          & Unit\\
        \hline
        $P_{DG,i}(t)$                  & \ac{dg} power set-point of the $i$-th generator                                                                                      & \SI{}{MW}\\
        $\delta_{m,i}(t)$              & \begin{tabular}{@{}l@{}}$m$-th interval of the \ac{sfoc} linearization of the \\ $i$-th generator\end{tabular}                       & \SI{}{MW}\\
        $z_i(t)$                       & \begin{tabular}{@{}l@{}}binary variable that identify the state of \\ the $i$-th \ac{dg}, on or off\end{tabular}                     & -\\
        $u_i(t)$                       & \begin{tabular}{@{}l@{}}binary variable that identify if $i$-th generator is \\ turning at time t\end{tabular}                       & -\\
        $P_{b}^{c}(t)$                 & \ac{bess} charging power set-point                                                                                                   & \SI{}{MW}\\
        $P_{b}^{d}(t)$                 & \ac{bess} discharging power set-point                                                                                                & \SI{}{MW}\\
        $SOC(t)$                       & \ac{bess} state of charge                                                                                                            & \%\\
        $s_j(t)$                       & \begin{tabular}{@{}l@{}}binary variable that identify which combination \\ is active\end{tabular}                                    & -\\
        $f_j(t)$                       & \begin{tabular}{@{}l@{}}integer variable that identify the number of \ac{dg}s \\ on in the $j$-th combination\end{tabular}           & -\\
        $z_b^c(t)$                     & \begin{tabular}{@{}l@{}}binary variable that identify if the \ac{bess} \\ is charging or not\end{tabular}                            & -\\
        $z_b^d(t)$                     & \begin{tabular}{@{}l@{}}binary variable that identify if the \ac{bess} \\ is discharging or not\end{tabular}                         & -\\
        $z_b(t)$                       & \begin{tabular}{@{}l@{}}binary variable that identify the state of \\ the \ac{bess}, on or off\end{tabular}                          & -\\
        \hline \hline
        \end{tabular}
\end{table}

\section{Simulations and Results Analysis} \label{SimAndRes}
The proposed algorithm is validated through the simulation of a shipboard microgrid made up of four \acp{dg} and a \ac{bess} (see Fig.\ref{fig:SystemArchitecture}). The rated parameters of each power source are collected in Table~\ref{tab:SimParam}.

Since the ship electrical load profile was not available, it has been modeled simulating a generic operating profile. The total ship load is composed by the hotel load, derived from the \ac{epla}, and the propulsive load, which depends on the \ac{sog}, as in \eqref{eq:Pload}.
\begin{equation}
   P_{load}(t) =
    \begin{aligned}[t]
    & P_{prop}\bigl(SOG(t,OC(t)\bigl) + \\ 
    & + P_{hotel}\bigl(OC(t)\bigl) \quad \forall \; t = 1:T\label{eq:Pload}
    \end{aligned}
\end{equation}

The \ac{sog} is simulated through a \ac{mc} model derived from real data of similar ships, as in \cite{MCGABRI}.
The hotel load $P_{hotel}$ is obtained from the \ac{epla} including a Gaussian noise with a variance of 5\%.

\begin{figure}[h!]
	\centering
        \includegraphics[width=7.4cm]{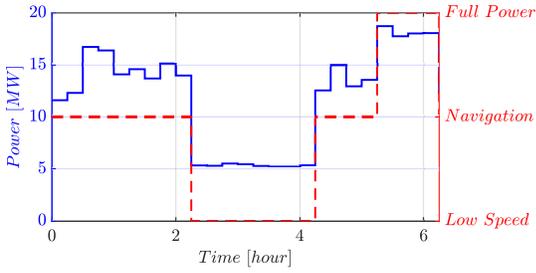}
	\caption{\ac{oc} and \ac{sog} profile.}\label{fig:Power_OCplot}
\end{figure}

Figure~\ref{fig:Power_OCplot} shows the load profile obtained through the above mentioned methodology. In this case the granularity of the load modelling is equal to \SI{15}{\min} and $T = \SI{6}{\hour}$. 

The simulation parameters are listed in Table~\ref{tab:SimParam}.
The parameter $v(t)$ is equal to one during the low speed and the navigation \ac{oc}. 
\begin{table}[h]
		\caption{Parameters of the simulations.}
		\label{tab:SimParam}
		\renewcommand{\arraystretch}{1.4}
        \begin{tabular}{cll}
        \hline \hline
        Parameter                & Value                                    & Description\\
        \hline
        $P_{DG,i = 1,2}^n$       & \SI{5.04}{\MW}                           & Rated power of i = 1,2 \ac{dg}\\
        $P_{DG,i = 3,4}^n$       & \SI{6.72}{\MW}                           & Rated power of i = 3,4 \ac{dg}\\
        $t_{min_u,i}$            & \SI{60}{\min}                            & Min up-time of i-th \ac{dg}\\
        $t_{min_d,i}$            & \SI{15}{\min}                            & Min down-time of i-th \ac{dg}\\
        $c_{DG,min}$             & 0                                        & Min power of \ac{dg}\\
        $c_{DG,max}$             & 1                                        & Max power of \ac{dg}\\
        $\Delta P_{r_u,i}$       & \SI[per-mode=symbol]{10}{\MW\per\min}    & Max \ac{dg} power ramp-up limit\\
        $\Delta P_{r_d,i}$       & \SI[per-mode=symbol]{10}{\MW\per\min}    & Max \ac{dg} power ramp-down limit\\
        $\alpha_{DG}$            & 1.1                                      & Max \ac{dg} overload in emergency\\
        $\beta_{DG}$             & 0.33                                     & Max \ac{dg} step in emergency\\
        $c_i$                    & \SI{200}{\EUR}                           & \ac{dg} start-up cost\\
        $c_f$                    & \SI[per-mode=symbol]{684}{\EUR\per\tonne}& fuel cost\\
        \hline
        $P_{b}^{n}$              & \SI{5}{\MW}                              & \ac{bess} nominal power\\
        $E_{n}$                  & \SI{5}{\MWh}                             & \ac{bess} nominal energy\\
        $SOC(1)$                 & 50\%                                     & Initial \ac{soc} of the battery\\
        $SOC(T)$                 & 50\%                                     & Final \ac{soc} of the battery\\
        $SOC_{max}$              & 80\%                                     & Max \ac{soc} of the battery\\
        $SOC_{min}$              & 20\%                                     & Min \ac{soc} of the battery\\
        $\eta_d$                 & 92\%                                     & \ac{bess} discharge efficiency\\
        $\eta_c$                 & 95\%                                     & \ac{bess} charge efficiency\\
        $c_{c,min}$              & 0                                        & Min charging C-rate\\
        $c_{c,max}$              & 1                                        & Max charging C-rate\\
        $c_{d,min}$              & 0                                        & Min discharging C-rate\\
        $c_{d,max}$              & 2                                        & Max discharging C-rate\\
        $\alpha_{b}$             & 3                                        & Max \ac{bess} overload in emergency\\
        \hline
        $M$                      & $10^9$                                   & Big M parameter\\
        $N$                      & 4                                        & Number of \ac{dg}\\
        $n_i$                    & 10                                       & Number of \ac{sfoc} intervals curve\\
        \hline \hline
        \end{tabular}
\end{table}

\begin{figure}[h]
	\centering
        \includegraphics[width=\columnwidth]{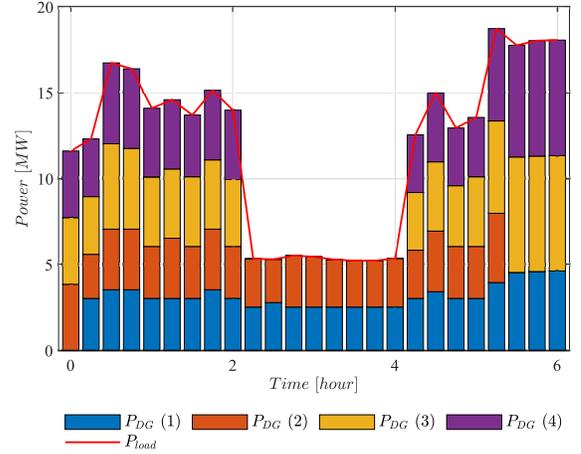}
	\caption{Simulation results, SC1, only \acp{dg}.} \label{fig:6h_DG}
\end{figure}

\begin{figure}[h]
	\centering
        \includegraphics[width=\columnwidth]{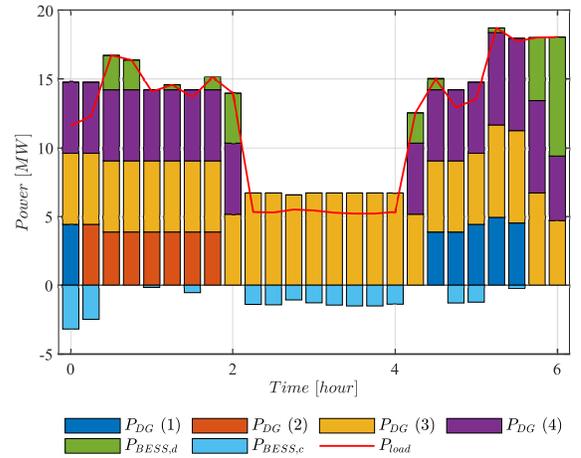}
	\caption{Simulation results, SC2, \acp{dg} and \ac{bess}.}\label{fig:6h_BESS_and_DG}
\end{figure}

The proposed power management strategy is tested on two study cases. In the first study case (SC1) the \ac{bess} is not considered, while in the second study case (SC2) the \ac{bess} is active. Figure~\ref{fig:6h_DG} shows the results of  the optimization for the SC1, wherein the power generation is provided only by the \ac{dg}s. Figure~\ref{fig:6h_BESS_and_DG} shows the results obtained for the SC2, where the \ac{bess} is available.

The results are reported in Table~\ref{tab:Results}, which summarizes the fuel cost, the amount of fuel required, the total $CO_2$ emission, and the average loading factor of the \acp{dg} ($LF_{avg}$).
\begin{table}[H]
    \centering
    \caption{Simulations results.}
    \label{tab:Results}
    {
    \begin{tabular}{cccc}
        \midrule
        \midrule
        \multicolumn{4}{c}{SC1 - only \acp{dg}}\\
        \midrule
        Total Cost            & Total Fuel      & CO$_2$        & $LF_{avg}$\\
        $[\SI{}{\EUR}]$       & $[kg]$          & $[kg]$        & [\%]\\
        \midrule
        8792                  & 12710           & 40.74         & 63\\
        \midrule
        \midrule
        \multicolumn{4}{c}{SC2 - \acp{dg} and \ac{bess}}\\
        \midrule
        Total Cost            & Total Fuel      & CO$_2$        & $LF_{avg}$\\
        $[\SI{}{\EUR}]$       & $[kg]$          & $[kg]$        & [\%]\\
        \midrule
        8672                  & 12520           & 40.13         & 79\\
        \midrule
        \midrule
    \end{tabular}
    }
\end{table}

The fuel costs are obtained considering the usage of the \ac{vlsfo} at a cost per metric tonne of \SI[per-mode=symbol]{684}{\EUR\per\tonne}, obtained from \cite{FuelPrice}. In addition, each start-up has an associated cost of \SI{200}{\EUR}, as previously reported (see \ref{tab:SimParam}).
The total $CO_2$ emissions have been calculated considering an emission factor of \SI{3.206}{\kg} of $CO_2$ per metric tonne \cite{IMOco2}.

The comparison of these two study cases shows that the \ac{bess} allows to reduce the fuel consumption by 1.5\% (\SI{190}{\kg} of fuel saving with respect to the case without \ac{bess}). As a consequence, also the $CO_2$ and the total cost are reduced.
In the SC2 the $LF_{avg}$ is much closer to point of lowest specific consumption, which is approximately equal to 80\% of the nominal power. It is important to highlight that both the study cases exploit the optimal control strategy.

\begin{figure}[h]
	\centering
        \includegraphics[width=8cm]{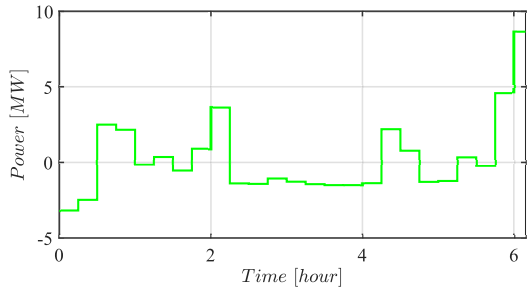}
	\caption{\ac{bess} power.}\label{fig:BESSpower}
\end{figure}

\begin{figure}[h]
	\centering
        \includegraphics[width=8cm]{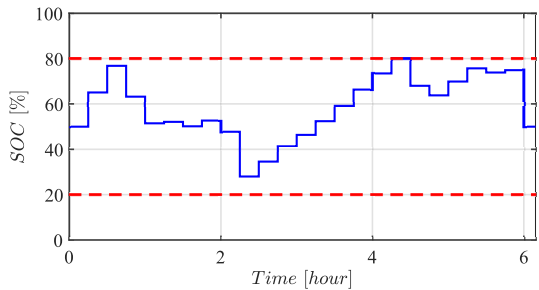}
	\caption{Battery \ac{soc}.}\label{fig:SOCplot}
\end{figure}

Figures~\ref{fig:BESSpower} and \ref{fig:SOCplot} illustrate \ac{bess} power and \ac{soc}, respectively. It is worth noting that the \ac{soc} is within the selected thresholds, and that the final values matches the initial requirements.

\section{Conclusions}
This work proposes an optimal power management strategy for shipboard microgrids equipped with \acp{dg} and a \ac{bess}. The optimization is aimed at optimally dispatching the power, in order to ensure reliable power
supply at minimum cost and with minimum environmental impact.

The algorithm is validated through the simulation of an all electric ship equipped with four \acp{dg} and a \ac{bess}. A comparison between two study cases have been conducted: the results show the that the presence of the \ac{bess} helps significantly the system to be more efficient, with savings around 1.5\% of fuel (\SI{190}{\kg}) in 6 hours.

Future developments will be devoted to the implementation of a model predictive control algorithm with load forecast capabilities.

\bibliographystyle{IEEEtran}
\bibliography{biblio}

\begin{thebibliography}{10}
\providecommand{\url}[1]{#1}
\csname url@samestyle\endcsname
\providecommand{\newblock}{\relax}
\providecommand{\bibinfo}[2]{#2}
\providecommand{\BIBentrySTDinterwordspacing}{\spaceskip=0pt\relax}
\providecommand{\BIBentryALTinterwordstretchfactor}{4}
\providecommand{\BIBentryALTinterwordspacing}{\spaceskip=\fontdimen2\font plus
\BIBentryALTinterwordstretchfactor\fontdimen3\font minus
  \fontdimen4\font\relax}
\providecommand{\BIBforeignlanguage}[2]{{%
\expandafter\ifx\csname l@#1\endcsname\relax
\typeout{** WARNING: IEEEtran.bst: No hyphenation pattern has been}%
\typeout{** loaded for the language `#1'. Using the pattern for}%
\typeout{** the default language instead.}%
\else
\language=\csname l@#1\endcsname
\fi
#2}}
\providecommand{\BIBdecl}{\relax}
\BIBdecl

\bibitem{MUELLER2023114460}
N.~Mueller, M.~Westerby, and M.~Nieuwenhuijsen, ``Health impact assessments of
  shipping and port-sourced air pollution on a global scale: A scoping
  literature review,'' \emph{Environmental Research}, vol. 216, p. 114460,
  2023.

\bibitem{MARPOL}
\emph{Amendments to the annex of the protocol of 1978 relating to the
  international convention for the prevention of pollution from ships, 1973},
  IMO Convention MEPC 116(51), April 2004.

\bibitem{fit455}
E.~Commission, ``Fit for 55—delivering the {EU}’s 2030 climate target on
  the way to climate neutrality,'' Jul 2021.

\bibitem{BESSuse}
J.~F. Hansen and F.~Wendt, ``History and state of the art in commercial
  electric ship propulsion, integrated power systems, and future trends,''
  \emph{Proceedings of the IEEE}, vol. 103, no.~12, pp. 2229--2242, 2015.

\bibitem{IACS}
\emph{SC1 Main source of electrical power - Interpretations of the
  International Convention for the Safety of Life at Sea {(SOLAS)}, 1974 and
  its Amendments}, International Association of Classification Societies
  {(IACS)} Std., 1974, Rev.2, Feb. 2021.

\bibitem{RadanBlackout}
D.~Radan, T.~Johansen, A.~Sørensen, and A.~Adnanes, ``Optimization of load
  dependent start tables in marine power management systems with blackout
  prevention,'' vol.~4, 12 2005.

\bibitem{twostepPMS}
S.~Fang, Y.~Xu, Z.~Li, T.~Zhao, and H.~Wang, ``Two-step multi-objective
  management of hybrid energy storage system in all-electric ship microgrids,''
  \emph{IEEE Transactions on Vehicular Technology}, vol.~68, no.~4, pp.
  3361--3373, 2019.

\bibitem{SecurityUC}
S.~Mashayekh and K.~L. Butler-Purry, ``Security constrained power management
  system for the ng ips ships,'' in \emph{North American Power Symposium 2010},
  2010.

\bibitem{kanellos}
F.~D. Kanellos, ``Optimal power management with ghg emissions limitation in
  all-electric ship power systems comprising energy storage systems,''
  \emph{IEEE Transactions on Power Systems}, vol.~29, no.~1, pp. 330--339,
  2014.

\bibitem{EPLAiees}
A.~Boveri, F.~D’Agostino, P.~Gualeni, D.~Neroni, and F.~Silvestro, ``A
  stochastic approach to shipboard electric loads power modeling and
  simulation,'' 2018, pp. 1--6.

\bibitem{Doerry}
N.~Doerry, ``Electric power load analysis,'' \emph{Naval Engineers Journal},
  vol. 124, pp. 45--48, 12 2012.

\bibitem{wartsila46f}
\emph{Wärtsilä 46F, product guide}, Wärtsilä, 2020, dAAB605814.

\bibitem{SFOCparbolic}
P.~Ghimire, M.~Zadeh, J.~Thorstensen, and E.~Pedersen, ``Data-driven efficiency
  modeling and analysis of all-electric ship powertrain: A comparison of power
  system architectures,'' \emph{IEEE Transactions on Transportation
  Electrification}, vol.~8, no.~2, pp. 1930--1943, 2022.

\bibitem{SOCdef}
K.~Hein, Y.~Xu, G.~Wilson, and A.~K. Gupta, ``Coordinated optimal voyage
  planning and energy management of all-electric ship with hybrid energy
  storage system,'' \emph{IEEE Transactions on Power Systems}, vol.~36, no.~3,
  pp. 2355--2365, 2021.

\bibitem{PIECEWISELIN}
M.~Carrion and J.~Arroyo, ``A computationally efficient mixed-integer linear
  formulation for the thermal unit commitment problem,'' \emph{IEEE
  Transactions on Power Systems}, vol.~21, pp. 1371 -- 1378, 09 2006.

\bibitem{MCGABRI}
S.~Massucco, G.~Mosaico, M.~Saviozzi, F.~Silvestro, A.~Fidigatti, and
  E.~Ragaini, ``An instantaneous growing stream clustering algorithm for
  probabilistic load modeling/profiling,'' in \emph{2020 International
  Conference on Probabilistic Methods Applied to Power Systems (PMAPS)}, 2020.

\bibitem{FuelPrice}
{Ship and Bunker}, ``{Average Bunker Prices},''
  \url{https://shipandbunker.com/prices/av}, 2023.

\bibitem{IMOco2}
{International Maritime Organization {(IMO)}}, ``Fourth greenhouse gas study
  2020,'' \url{www.imo.org}, 2021.

\end{thebibliography}

\end{document}